# Unveiling spectral purity and tunability of terahertz quantum cascade laser sources based on intra-cavity difference frequency generation

Luigi Consolino,[1†] Seungyong Jung,[2†] Annamaria Campa,[1] Michele De Regis,[1] Shovon Pal,[3] Jae Hyun Kim[2], Kazuue Fujita[4], Akio Ito[4], Masahiro Hitaka[4], Saverio Bartalini,[1] Paolo De Natale,[1] Mikhail A. Belkin[2††] and Miriam Serena Vitiello [3*]

[1]CNR- Istituto Nazionale di Ottica and LENS (European Laboratory for Non-linear Spectroscopy ), Via Carrara 1, 50019 Sesto Fiorentino (FI), Italy

[2]Department of Electrical and Computer Engineering, The University of Texas at Austin, Austin, TX 78712

[3]NEST, CNR - Istituto Nanoscienze and Scuola Normale Superiore, Piazza San Silvestro 12, 56127, Pisa, Italy

[4]Central Research Laboratory, Hamamatsu Photonics K.K. Hamakitaku, Hamamatsu, 434-8601, Japan

\* miriam.vitiello@sns.it
[††]mbelkin@ece.utexas.edu

† These authors contributed equally to this work.

## Abstract

Terahertz sources based on intra-cavity difference-frequency generation in mid-infrared quantum cascade lasers (THz DFG-QCLs) have recently emerged as the first monolithic electrically-pumped semiconductor sources capable of operating at room-temperature (RT) across the 1-6 THz range. Despite tremendous progress in power output, that now exceeds 1 mW in pulsed and 10 μW in continuous-wave regime at room-temperature, knowledge of the major figure of merits of these devices for high precision spectroscopy, such as spectral purity and absolute frequency tunability, is still lacking. Here, by exploiting a metrological grade system comprising a terahertz frequency comb synthesizer, we measure, for the first time, the free-running emission linewidth (LW), the tuning characteristics, and the absolute frequency of individual emission lines of these sources with an uncertainty of 4 x $10^{-10}$. The unveiled emission LW (400 kHz at 1ms integration time) indicates that DFG-QCLs are well suited to operate as local oscillators and to be used for a variety of metrological, spectroscopic, communication, and imaging applications requiring narrow-linewidth THz sources.

## Introduction

After two decades from their invention, [1] quantum cascade lasers (QCLs) have reached impressive performance levels in the mid-infrared (mid-IR, λ≈2.5-30 μm) and terahertz (THz, 0.3-10 THz) spectral regions. Mid-IR QCLs can now operate continuous-wave (CW) at room-temperature with multi-watt output power and high wall plug efficiencies exceeding 25%.[2,3] THz QCL heterostructures embedded in single-plasmon or double-metal waveguides have demonstrated lasing in the 1.2-5.4 THz frequency range,[4] with peak output powers that can reach watt-levels.[5,6] CW THz QCLs have established themselves as versatile narrow-linewidth (LW) sources with applications in high-resolution spectroscopy, gas sensing, and heterodyne detection. To date, cryogenically-cooled THz QCLs have shown high spectral purity with intrinsic LWs as low as ~ 100 Hz.[7] Their typical free-running LWs are dominated by excess frequency noise, leading to broadening up to several hundred kHz.[8-10] If frequency or phase stabilized, these sources have achieved fast LWs of 30-60 kHz, 20 kHz and 6.3 kHz, [8-13] depending on the quality of the reference and of the locking set-up. Narrow LW THz QCL sources are highly desired for a number of applications,[14] including heterodyne detection for far-infrared astronomy, molecular frequency metrology,[15] high-resolution coherent imaging and telecommunications, where they can provide the carrier wave for broadband wireless links.[16]

Despite the tremendous progress, the onset of temperature-activated LO phonon scattering of electrons in the upper laser state, combined with the difficulty of achieving selective electron injection into the closely spaced laser states, [17] still limits operation of the best THz QCLs to temperatures of ~ 200 K in pulsed mode [18] and slightly below 130 K in CW operation.[19] Room-temperature (RT) operation is, however, highly desired for applications and largely simplifies any experimental set-up.

To address the need for RT THz sources, an alternative approach has been implemented based on intra-cavity difference-frequency generation (DFG) in mid-IR QCLs.[20-22] With the introduction of a Cherenkov-effect scheme for THz extraction,[23,24] THz DFG–QCLs have made dramatic progress in the last five years.[25] This foundational approach has several appealing aspects: since the intersubband transitions within the QCL material itself provide the $\chi^{(2)}$ nonlinearity, the sources are monolithic and can be operated at RT, similarly to other mid-IR QCLs; additionally, wide THz tunability can be obtained with only modest tuning of the mid-IR modes. [26,27] Depending on the mid-IR pump spacing, THz emission can be varied in the entire 1–6 THz range and beyond, [28] with limitations set only by the material losses and reduction of THz DFG efficiency.[29]

## Results

**THz DFG QCL design and characterization.**

In spite of the mentioned major advances, the THz emission of DFG-QCLs has so far only been characterized using low resolution (~4 GHz) Fourier-transform infrared spectrometers (FTIRs) and their emission LW is still unknown. Here we report LW measurement of a free-running single frequency THz DFG-QCL. By beating the free-running DFG-QCL with a free-space THz frequency comb synthesizer (FCS), we retrieve an upper limit for the DFG-QCL emission LW of 125 kHz in 20 μs and 400 KHz at 1ms, at an emission frequency of 2.58 THz. Our work establishes THz DFG-QCLs as a viable alternative to THz QCLs for numerous applications that require narrow-LW emitters.

Details of the laser structure are given in Methods Section. The active region of the devices employed in this work was made up of two stacks of 20 $In_{0.53}Ga_{0.47}As/In_{0.52}Al_{0.48}As$ mid-IR QCL stages. The stages, based on the bound-to-continuum active region design,[30,31] were designed to provide peak gain at $\lambda_1$ = 8.5 μm and $\lambda_2$ = 9.5 μm. The laser bandstructure was designed to possess giant second-order optical nonlinearity, associated with intersubband transitions. The estimated nonlinearities for a THz DFG at 2.5 THz are $|\chi^{(2)}|$=9.3 nm/V and $|\chi^{(2)}|$=5.4 nm/V for the short- and long-wavelength active regions, respectively. Devices were processed into 2-mm-long 12-μm-wide buried-heterostructure lasers with dual-frequency mid-IR distributed feedback (DFB) grating along the laser waveguide. The DFB grating was designed to select mid-IR pumps wavelengths at approximately 8.47 μm and 9.11 μm in order to produce THz DFG emission at approximately 2.5 THz, within the optical bandwidth of the hot electron bolometer (HEB) used in the experiments. The front facet of the laser bar substrate was polished at 30° for out-coupling of the THz Cherenkov DFG emission,[25] while the back device facet was covered with a high-reflective coating with $Al_2O_3$/Ti/Au=120nm/10nm/60nm. The laser bars were then mounted in an epi-up configuration on a copper block and cooled in open-loop $LN_2$ cryostats for the initial characterization.

Figure 1 shows the measured emission data for a 2-mm-long 12-μm-wide buried-heterostructure waveguide Cherenkov THz DFG-QCLs operated in CW mode at T=65 K and T = 85 K, which we selected for this study. Figure 1(a) shows the FTIR spectra, obtained in rapid-scan mode with a spectral resolution of 0.125 cm$^{-1}$ (3.75 GHz), with the two mid-IR pumps and the THz DFG emission from the device. This device emitted a single THz frequency centered at approximately 2.58 THz at low pump currents; at pump currents above 450 mA, a second THz DFG emission line at 2.49 THz appeared in the spectra. A red shift in the frequency of the mid-IR pumps and for the THz difference-frequency is observed as the bias current is increased. This dependence is attributed to an increase of the active region temperature as the pump current increases. Light-current and voltage-current characteristics of mid-IR and THz outputs are shown in Fig. 1(b) and 1(c), respectively.

The temperature tuning and the thermal resistance of our device are determined by measuring the emission spectra in pulsed mode (400 ns pulses at 25 kHz repetition frequency) at heat-sink temperatures ($T_H$) varied from 45 K to 200 K. Under this pulsed operational regime, we can safely assume that the active region lattice temperature is nearly coincident with the heat-sink one. The spectral positions of the mid-IR pumps and the THz difference-frequency as a function of $T_H$ are given in Fig. 2a and 2b, respectively. By comparing the spectral positions of the mid-IR pumps in Fig. 1(a) and Fig. 2(a), we can retrieve the thermal resistance of our device, which is approximately 6.6 K/W at 45 K and 7.9 K/W at 85 K.

**Detection of the beating signal with a THz FCS and LW characterization.**

To probe the LW of the THz emission from the device, we investigated the beat-note signal arising from the beating between the THz emission from the DFG-QCL and the free space THz FCS. The experimental set-up is shown in Fig. 3(a).

The THz frequency comb is generated in a MgO-doped lithium niobate waveguide by optical rectification in the Cherenkov configuration using a femtosecond mode-locked fiber laser emitting around 1.5 μm at a repetition rate ($f_{rep}$) adjustable around 250 MHz. The optical rectification process produces a zero-offset free-space THz comb with a spectral content broader than the THz QCL tunability range.[11] The frequency ($f_n$) of each tooth of the THz FCS can be parameterized as:

$$f_n = n * f_{rep} \qquad (1)$$

where *n* is the comb tooth order.

Both the THz DFG-QCL and the reference THz comb beams were collected and collimated by a set of 90° off-axis parabolic mirrors with an equivalent focal length of 25.4 mm. The two beams are then combined on a mylar film which acts as a THz beam splitter and are then sent to the HEB, (Scontel model RS 0.3–3T-1) with a 250 MHz electrical bandwidth. The electrical signal from the HEB is sent to a fast Fourier transform real-time spectrum analyzer (Tektronix RSA5106A). Within the HEB bandwidth, it is possible to detect both the signal at 250 MHz, generated by the intermodal beating of the THz comb, and the pair of beat-notes generated by heterodyning the THz DFG QCL with the closest right and left comb teeth.

Figure 3(b) shows one of these beat note signals. In this case, the beating is with the right (higher-frequency) comb tooth, and the beat-note frequency ($f_b$) is given by:

$$f_b = Nf_{rep} - f_{QCL} \qquad (2)$$

where $f_{QCL}$ is the THz DFG-QCL frequency and N is the order of the closest THz FCS tooth involved. The beat-note spectrum in Fig. 3(b) can be described by a Gaussian function, whose profile has been used to fit each line-shape. Since the LW of the THz comb tooth involved in the beating process (~130 Hz @ 1 s, as experimentally demonstrated)[11,15] is negligible with respect to the DFG-QCL THz emission LW, the full width at half maximum (FWHM) of the Gaussian profiles provides an accurate quantitative estimation of the THz emission LW of our device.

The LW of the 2.58 THz emission line of the CW-operated device has been measured as a function of the observation time $t$ at two different heat sink temperatures (45 K and 78 K) as shown in Fig. 4(a). The range of analysis is limited at short timescales at 20 µs, corresponding to the acquisition time of the single FFT spectrum for which the beat note width starts to be limited by resolution bandwidth of the instrument (100 kHz, in this case). At this timescale we measure an upper limit of the QCL LW of 125 kHz.

The logarithmic dependence of the LW with $t$, clearly indicates the presence of a 1/f noise component in the frequency noise spectrum between 1 Hz and 1 kHz that likely comes from the typical "pink" frequency-noise of the mid-IR pumps of our DFG-QCLs.[32,33] This is also confirmed by the noise decrease at increasing temperatures (see Fig.4(a)), in agreement with previous experiments on the 1/f noise of mid-IR QCLs [33-35] that showed a similar trend.

A quantitative comparison with previous LW measurement reports of mid-IR QCLs can help to understand the possible correlations between the emission LW of the mid-IR pumps and the THz emission LW. To this purpose, we applied the inverse method proposed in Ref. 36 to the data in Fig. 4(a) in order to retrieve the frequency noise power spectral density (FNPSD) of our devices (see Methods Section for details). The FNPSD of THz DFG-QCLs can be directly compared with the FNPSDs of a THz QCL operating at 47.5 K [7] and of a mid-IR QCL operating at 85 K, [32] as shown in Fig. 4 (b). The plot clearly unveils that the FNPSD of the DFG-QCL, even in the upper edge of the investigated temperature range ($T_H$ = 45 K-85 K), is almost one order of magnitude lower than that of the mid-IR QCL. The reduction on the emission LW of THz DFG can be explained by the correlation between the phase/frequency noises of the two mid-IR pumps of our device that are partly compensated by the DFG process. On the other hand, the FNPSD of a single- mode bound-to-continuum THz QCL, exploiting a single plasmon waveguide, shows a significantly narrower LW (~ a factor of seven). Nevertheless, the reported measurements indicate that the LW of THz DFG-QCLs is already suitable for heterodyne THz detection, and we expect that it can be further reduced with frequency/phase stabilization. It is worth noting that no significant dependence of the LW on the driving current can be evidenced.

**Absolute emission frequency and tuning coefficients.**

Equation (2) allows us to determine the absolute frequency position of the two THz emission peaks in the spectra in Fig. 1 and to perform high-resolution studies of the dependence of the emission frequency on the DFG-QCL operating conditions. First, the absolute value of the emitted frequencies is measured by tuning the repetition rate $f_{rep}$ of the THz FCS pump laser from 249 MHz to 251 MHz, and by observing the linear dependence of $f_b$ on $f_{rep}$. By fitting the two different data sets with Eq. (2), as shown in Fig. 5(a), we determined the absolute frequencies of the two modes. The retrieved values are $\nu_1$ = 2581470(184) MHz and $\nu_2$ = 2493084(64) MHz for the first and second THz mode, respectively, in agreement with the measured FTIR spectra (Figs.1a-b). It is worth noting that the obtained uncertainties are much smaller than the spacing of the THz comb teeth, and this allows to univocally determine the order N of the comb teeth that are beating with the two modes, as confirmed by the errors on N that are smaller than unity. Once N is determined from the measurement of the beat-note frequency, it is possible to retrieve back the instantaneous absolute frequency of the free-running QCL with a 1 kHz uncertainty and a resulting relative accuracy of about $4 \times 10^{-10}$.

Figure 5(b) shows the absolute values of the THz emission frequency of the first mode as a function of $T_H$ for the CW operation with 500 mA pump current. Measurements were performed while varying $T_H$ from 45 K to 85 K, which corresponds to QCL waveguide core lattice temperatures varying from ~ 90 K to 140 K, using the device thermal resistances calculated earlier. The observed temperature dependence of the THz spectral position is nearly linear with a slope coefficient of about -193 MHz/K. This value is in agreement with the THz frequency dependence deduced from the data in Fig. 2(b) for the same range of laser core temperatures.

Figure 5(c) shows the dependence of the frequency of the first THz mode on device pump current. Surprisingly, for the device operating at $T_H$ = 45 K, the THz emission frequency initially blue-shifts with a tuning rate of about +3.3 ± 0.2 MHz/mA, when the injection current is slightly above the laser threshold. A turning point is visible at ~ 450 mA, and then the DFG QCL frequency red-shifts with a tuning coefficient of approximately -2.16 ± 0.07 MHz/mA. At 65 K, the inversion point is shifted towards lower current. By further increasing $T_H$ to 85 K, the data show a monotonic decrease in THz frequency with pump current with a slope of approximately -6 MHz/mA.

The observed dependence of the THz frequency on pump current cannot be explained by considering only the Ohmic heating of the device core, which would predict a monotonic decrease in the THz frequency with a slope of approximately -17 − -20 MHz/mA. The value is obtained using the measured temperature tuning coefficient of -193 MHz/K, the thermal resistance of 6.6−7.9 K/W deduced earlier, and the I-V characteristic of the device shown in Fig. 1. The observed trend is likely related to the gain dynamics behind the active region architecture. In semiconductor lasers, the refractive index variations (and corresponding emission frequency variations) due to the gain change with pump current can be estimated via the LW enhancement factor ($\alpha_e$)[37]. Unlike diode lasers, the LW enhancement factor of QCLs can be either positive or negative, depending on the position of the emission frequency relative to the gain peak and the dynamics of the gain spectrum.[38-41] Although no correlation can be found with the measured FTIR spectra due to the inherent resolution limits, we can tentatively ascribe the different tuning trends to the different values of the $\alpha_e$ for the two mid-IR pumps. Further investigation of this effect will be performed in the future by combining high-resolution mid-IR and THz spectral measurements.

**Discussion**

In conclusion, by assessing the free-running LW, the tuning characteristics and the absolute frequency of a THz DFG-QCL, via a high-resolution heterodyne technique, we have demonstrated, for the first time, that CW THz DFG-QCLs are suitable as metrological tools for high-resolution spectroscopy at THz frequencies and as local oscillators for heterodyne detection. Furthermore, the accuracy of the used technique allowed unveiling the current and temperature-induced tuning, providing relevant information to build low-chirp and low frequency-noise THz emitters by novel design of the gain media for these devices.

**Materials and Methods**

### Device design

The laser structure was grown on a 350-μm thick semi-insulating InP substrate by a metal organic vapor phase epitaxy system. The growth started with the 400-nm-thick current injection layer doped to $1\times10^{18}$ cm$^{-3}$, followed by the 5-μm-thick InP lower cladding layer doped to $1.5\times10^{16}$ cm$^{-3}$, 200-nm-thick In$_{0.53}$Ga$_{0.47}$As waveguide layer, 2.6-μm-thick active region made of two 20-period stacks of quantum cascade stages, and 300-nm-thick In$_{0.53}$Ga$_{0.47}$As DFB grating layer. Both In$_{0.53}$Ga$_{0.47}$As layers were doped to $1.5\times10^{16}$ cm$^{-3}$. The layer sequence of the short-wavelength stage (bottom stack), starting from the injection barrier, is **4.1**/1.8/**0.7**/5.5/**0.9**/5.3/**1.1**/4.8/ **1.5**/3.7/**1.6**/ 3.5/**1.6**/3.3/**1.8**/3.1/**2.0**/2.9/**2.4**/<u>2.7</u>/**2.6**/<u>2.7</u>/**3.0**/2.7, where the layer thickness is given in nanometers. The layer sequence of the long-wavelength stage (top stack) is **3.8**/2.0/**0.9**/6.0/**0.9**/5.9/**1.0**/5.0/**1.1**/4.0/**1.5**/3.4/**1.5**/3.3/**1.6**/3.0/**1.9**/3.0/**2.3**/<u>3.1</u>/**2.5**/<u>3.2</u>/**2.9** /3.0. The bold and underlined fonts denote the barrier and doped (Si $=1.2\times10^{17}$ cm$^{-3}$) layers, respectively. A 350-nm-deep first-order mixed DFB grating structure [31] was then implemented in the DFB grating layer and (partly) the device active region using e-beam lithography and dry etching. The estimated DFB coupling coefficient was 22 cm$^{-1}$, for both mid-IR pumps. The 12-μm-wide ridges were defined via wet etching and processed into buried heterostructures by lateral overgrowth of a Fe-doped semi-insulating InP layer for surface planarization. The growth was finalized with the 5-μm-thick InP cladding layer doped to $1.5\times10^{16}$ cm$^{-3}$ and 15-nm-thick In$_{0.53}$Ga$_{0.47}$As contact layer doped to $1\times10^{19}$ cm$^{-3}$. We used the side-injection scheme for lateral current injection and electroplated 5-μm-thick gold on top of our devices for efficient heat extraction.

### Calculation of the 1/f noise level from the linewidth measurements

Reference 36 provides a simplified method for retrieving the LW of any laser source over a given observation time $t_0$ from its FNPSD ($S$(f)). According to this procedure, and assuming a monotonic trend for the FNPSD, the laser linewidth dv($t_0$) (FWHM) is recovered by a simple integration of $S$(*f*):

$$\delta v(\tau_0) = \sqrt{8ln(2) \int_{f_0}^{f_l} S(f)df} \qquad (3)$$

where the lower bound of the integral, $f_0$, is $1/\tau_0$ and the upper bound of the integral, $f_l$, is defined as the frequency at which the FNPSD crosses the so called b-separation line:

$$S(f_l) = \frac{8ln(2)}{\pi^2} \cdot f_l \qquad (4)$$

The procedure well approximates the true value only if the following condition is fulfilled:

$$S(f_0) > \frac{8ln(2)}{\pi^2} \cdot f_0 \qquad (5)$$

Assuming that, in the explored frequency range (10 Hz – 10 kHz), the FNPSD of the laser has a flicker-type origin (i.e., $S(f)=A/f$), Eq. 4 gives a logarithmic-type function that can be used to fit the experimental values dn(t) of the laser LW at different timescales (Fig. 4(a)), thus allowing us to determine the parameter A. Figure 4(b) shows the two flicker-type functions that reproduce, via Eq. (3), the experimental LWs of our device operated at 45, and 78 K, together with the two flicker-type functions that directly fit the experimental FNPSDs measured from a mid-IR QCL operating at 85 K [32] and a THz QCL operating at 47.5 K [7]. Figure 4(b) allows checking *a posteriori* the validity of the condition in Eq. (5). The crossing point with the b-separation line falls above 10 kHz, which validates the model assumptions for the explored frequency range of 10 Hz – 10 kHz.

## Acknowledgments

**Funding:** This work was partially supported by the European Research Council through the ERC grant 681379 (SPRINT) (M.S.V.) and the United States National Science Foundation grant ECCS-1408511 (M.A.B.). J.H.K. acknowledges the support from the 863 Program of the Republic of Korea grant number 2013AA014402. M.A.B. acknowledges support from Alexander von Humboldt Foundation Friedrich Wilhelm Bessel Research Award.

**General**: M.A.B. acknowledges helpful discussions with Prof. Markus-Christian Amann of the Technical University of Munich.

**Competing interests:** The authors declare no competing financial interests.

## Figures and Tables

**Figure 1. Emission spectra and power output of the DFG-QCL used in our study. (a-b)** FTIR emission spectra of the two mid-IR pumps and the resulting THz DFG, collected at a heat sink temperature of 65 K (a) and 85 K (b) in rapid scan mode with a 0.125 cm$^{-1}$ (3.75 GHz) spectral resolution, plotted as a function of the driving current. Identical colors in the mid-IR and THz spectra correspond to identical driving current conditions. **(c)** Current-voltage and light-current characteristics of the mid-IR pumps and THz DFG for the device operated continuous-wave at 85 K.

**Figure 2. Frequency tuning.** Dependence of mid-IR pump frequencies **(a)** and THz difference-frequency **(b)** of the THz DFG-QCL on the heat-sink temperature ($T_H$). The device was operated in pulsed mode as described in the text. The dashed line in panel (b) represent the frequency difference of the two mid-IR peaks as extracted from panel (a), the round symbols with error bars indicate the experimentally measured THz peak positions.

**Figure 3. Experimental setup. (a)** The schematic of the experimental setup: The THz frequency comb is generated in a MgO-doped lithium niobate waveguide by optical rectification with Cherenkov phase-matching of a femtosecond mode-locked fiber laser, working at 1.5 μm wavelength. The optical rectification process produces a zero-offset free-space THz comb with the repetition rate of approximately 250 MHz of the pump femtosecond laser. The THz frequency comb and the emission from the DFG-QCL were overlapped on the sensor element of the HEB with a 250 MHz electrical

bandwidth. The beat note signals of the THz DFG-QCL emission with the nearby frequency comb lines were analyzed by a FFT spectrum analyzer, having a 40 MHz real time bandwidth. **(b)** A typical beat note spectrum observed on a spectrum analyzer for 2 ms integration time.

**Figure 4. Linewidth measurements. (a)** The width of the beat note at different time scales measured at two different operating temperatures of the device. The use of a FFT spectrum analyzer allows retrieving the beat-note spectra over different integration times, and therefore to evaluate the THz DFG-QCL emission linewidth at different time scales (ranging from 20 μs to approximately 20 ms). **(b)** Reconstruction of the frequency noise power spectral density (FNPSD) of the THz DFG-QCL emission. The plots have been obtained inverting the approach developed in Ref. 36 as discussed in Methods section. This procedure allows retrieving the FNPSD for frequencies where it is larger than the β-line ($8/\pi^2 \ln(2) f$), dashed in the lower part of the graph. Given the smallest time scale of 20 ms of our setup, our measurements do not include frequencies higher than 50 kHz.

**Figure 5. Absolute frequency measurements (a)** Absolute frequency of the DFG-QCL plotted as a function of the THz frequency comb repetition rate, while keeping the QCL driving current at 540 mA and the heat sink temperature ($T_H$) fixed at 75 K. **(b)** Frequency tuning of the 2.58 THz emission line of the DFG-QCL as a function of the QCL heat-sink temperature, measured while keeping both the comb repetition-rate and the QCL driving current fixed. The 2.49 THz emission line shows a similar dependence. **(c)** QCL frequency tuning of the 2.58 THz emission line as a function of the driving current at different $T_H$, measured while keeping the comb repetition-rate and the QCL driving current fixed. The 2.49 THz emission line shows a similar dependence.